\documentclass{hep99}
\usepackage[dvips]{graphicx}
\begin{document}

\title{Recent results from Selex
}
\author{
M.~Iori,
\centerline{University of Rome "La Sapienza" and INFN, Rome, Italy}       
\vskip 0.50cm                                                                 
\centerline{on behalf for the SELEX Collaboration\dag}                     
\vskip 0.50cm      
}

\abstract{
The SELEX experiment (E781) is a 3-stage magnetic spectrometer for
   a high statistics study of hadroproduction of charm baryons out to
   large $x_{F}$ using 650 GeV $\Sigma ^{-}$,$\pi ^{-}$ and $p$ beams. 
    The main features
   of the spectrometer are: a high precision silicon vertex system; 
   powerful particle identification provided by TRD
   and RICH; forward $\Lambda _{s}$ decay spectrometer; and 3-stage 
   lead glass photon
   detector.
   Preliminary results on asymmetry for $\Lambda _{c}$  produced 
   by $\Sigma ^{-}$, $\pi ^{-}$ and $p$ beams at $x_{F}>0.2$ 
   and precise measurements of the $\Lambda _{c}$, $D ^{0}$, and 
   preliminary $D _{s}$ 
   lifetimes are presented.
} 

\maketitle

\fntext{\dag}{M.~Iori,$^{18}$ 
G.~Alkhazov,$^{11}$ A.G.~Atamantchouk,$^{11}$ 
M.Y.~Balatz,$^{8}$ 
N.F.~Bondar,$^{11}$ 
P.S.~Cooper,$^{5}$ 
L.J.~Dauwe,$^{17}$ 
G.V.~Davidenko,$^{8}$ 
U.~Dersch,$^{9}$ 
A.G.~Dolgolenko,$^{8}$
% ...
$^{1}$Ball State University, Muncie, Indiana 47306
$^{2}$Bogazici University, Bebek 80815 Istanbul, Turkey
{$^{3}$Carnegie--Mellon University, Pittsburgh, Pittsburgh 15213}
{$^{4}$Centro Brasileiro de Pesquisas F\'{\i}sicas, Rio de Janeiro,
                  Brazil}
{$^{5}$Fermi National Accelerator Laboratory, Batavia, 
                  Illinois 60510}
{$^{6}$Institute for High Energy Physics, Protvino, Russia}
{$^{7}$Institute of High Energy Physics, Beijing, PR China}
{$^{8}$Institute of Theoretical and Experimental Physics, 
                  Moscow, Russia}
{$^{9}$Max--Planck--Institut f\"ur Kernphysik, 69117 Heidelberg,
                  Germany}
{$^{10}$Moscow State University, Moscow, Russia}
{$^{11}$Petersburg Nuclear Physics Institute, St. Petersburg, 
                   Russia}
{$^{12}$Tel Aviv University, 69978 Ramat Aviv, Israel}
{$^{13}$Universidad Aut\'onoma de San Luis Potos\'{\i}, 
                   San Luis Potos\'{\i}, Mexico}
{$^{14}$Universidade Federal da Para\'{\i}ba, Para\'{\i}ba, Brazil}
{$^{15}$University of Bristol, Bristol BS8~1TL, United Kingdom}
{$^{16}$University of Iowa,  Iowa City, Iowa  52242}
{$^{17}$University of Michigan--Flint, Flint, Michigan 48502}
{$^{18}$University of Rome "La Sapienza" and INFN , Rome, Italy}
{$^{19}$University of S\~ao Paulo, S\~ao Paulo, Brazil}
{$^{20}$University of Trieste and INFN, Trieste, Italy}
}

\section{Introduction}

Charm physics explores QCD phenomenology in both perturbative 
and nonperturbative
regimes. Production dynamics studies test leading order (LO)
and next to leading order (NLO) perturbative QCD.
The present fixed target experiments on charm hadroproduction are in 
qualitative agreement 
with perturbative QCD
calculations, but quantitative deviations from QCD are observed.
More experimental data, using different incident hadrons ($\pi$, $p$ and
$\Sigma ^{-}$), may help to illuminate hadron-scale physics: color-drag,
fragmentation, and intrinsic $k _{t}$ effects.
Charm lifetime measurements test models based on $1/ M_{Q}$ QCD 
expansions.  Precise measurement of the lifetimes of charm meson and baryon weak decays
are also important to understand perturbative QCD in term of 
non-spectator W-annihilation as well Pauli interference. For example,
non-leptonic decay rate differences between W-exchange in $D^{0}$ and
$D^{\pm} _{s}$ may produce a lifetime difference of order $10-20 \%$ 
\cite{Frix97}.
  
\section{The Selex spectrometer}

The SELEX experiment at Fermilab is a 3-stage magnetic spectrometer.
The 600 GeV/$c$ Hyperon beam of negative polarity 
contains equal fraction of 
$\Sigma$ and $\pi$.  The positive 
beam is composed of 92\% of protons and the rest $\pi$'s.
Beam particles are identified by a Transition Radiation detector (BTRD).
The spectrometer was designed to study charm production in the forward 
hemisphere with good mass and decay vertex resolution for charm momentum
in a range of 100-500 GeV/$c$.
The vertex region is composed of 5 targets (2 Cu and 3 C). The total
target thickess is 5\% of $\lambda _{int}$ for protons and the targets are
separated by 1.5 cm. Downstream of the targets there are 20  
silicon planes with a strip pitch of 20-25 $\mu$m disposed in X,Y,U and
V views.
The M1 and M2 magnets effect a momentum cutoff of 2.5 GeV/$c$ and 15 GeV/$c$ 
respectively. A RICH detector, filled with Neon at room temperature and pressure, 
provides a single track ring radius resolution of 1.4\% and
 2$\sigma$ $K/ \pi$ separation up to about
165 GeV/$c$. A computational filter uses   tracks identified by the RICH 
and linked to the vertex silicon by the PWCs to make a full reconstruction
of the secondary vertex. Events consistent with only a primary vertex
are rejected.  A layout of the spectrometer can be found elsewhere \cite{spec}.

\section{Data set and charm selection}

The charm trigger is
very loose. It requires a valid beam track, two of opposite charge 
tracks to the beam with momentum $>15$ GeV/$c$, 
two high momentum
tracks linked to the Silicon vertex detector, and unconnected to 
all other tracks from the primary vertex. We triggered on about 1/3 of all
the inelastic interactions. About 1/8  of them are written on the tape
for a final sample of about $0.9\cdot 10^9$
 events.
In the analysis secondary vertices were reconstructed if the
$\chi ^{2}$ of all tracks was inconsistent with single primary vertex.
The RICH detector labelled all particles above 25 GeV/$c$.

All data reported here resulted from a first pass through
the data. 


\begin{figure}
\vspace{7cm}
\includegraphics{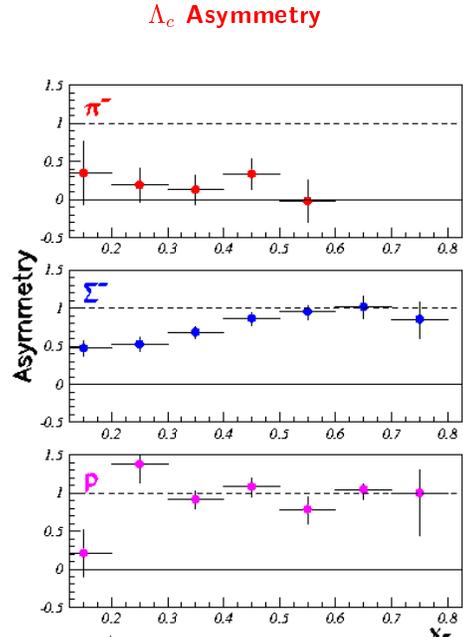}
\caption{ ${\Lambda _{c}}^{+}$ asymmetry versus $x _{F}$  for different beams.}
\end{figure}

\begin{figure}
\vspace{7cm}
\includegraphics{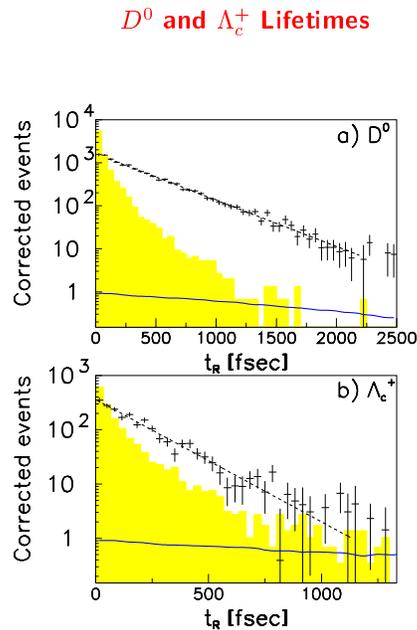}
\caption{The acceptance corrected reduced proper lifetime 
distribution for the background subtracted signal (points)
and the mass sideband (shaded) region for a) $ D ^{0}$
and b) ${\Lambda _{c}}$ lifetime. The dashed line is the lifetime fit.
The solid line is the acceptance as function of reduced proper time.}
\end{figure}


\begin{figure}
\vspace{7cm}
\includegraphics{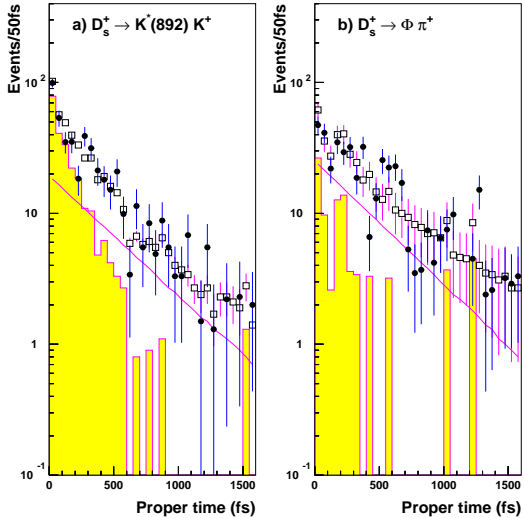}
\caption{Corrected reduced proper time distributions for events under the
$ D _{s}$ peak (dots) and results from the maximum likelihood fit
(squares). The hatched region shows the fitted background.
The dashed line shows the signal lifetime distribution.}
\end{figure}

\subsection{Charm performance}

The requirements to study charm physics and to reduce the
background are: 
\par
good decay vertex resolution, mass 
resolution and particle identification. 
\par
The vertex algorithm
provides an average longitudinal error, $\sigma _{z}$, on the primary
and secondary vertex of 170 $\mu$m and 600$\mu$m respectively. Their
combination, $\sigma$, is equal to 650 $\mu$m. In the  $\Lambda _{c}$
($D _{s}$) sample, the average momentum is 220 (200) GeV/$c$, 
corresponding to a time resolution of 20 (18) fs.
\par
To reduce background we only keep charm candidates that decay
after a longitudinal separation from the primary vertex of at least $z _{min}= 8 \sigma $.
\par
The RICH detector gives  a good
separation $\pi / K$ for track momentum of 100 GeV/$c$,
typical momenta  for charm decay tracks (p,K).
To measure the lifetime we require each track extrapolated to the primary 
vertex z position must miss by a transverse distance length 
$t \ge$ 2.5 times its error $\sigma _{t}$; the secondary
vertex must lie outside any charm target by at least 0.5 mm; 
decays must occur within a given fiducial region; and
the proton and kaon tracks are identified by RICH.

\section{Charm and anticharm asymmetry}

The data yields were corrected by the geometrical 
acceptance and reconstruction efficiency. The efficiency
difference for baryon and antibaryon is less than 3\%.  
\par
For both baryon beams  ($\Sigma ^{-} , p$) there is a large $x_{F}$
enhancement of ${\Lambda _{c}}^{+}$ production.
Fig.~1 shows the $\Lambda _{c} ^{+}$ asymmetry $x _{F}$ dependence for 
different beams. The hadroproduction asymmetry is defined as:
$A= ({\sigma _{c} - \sigma _{\overline{c}}})/ ({\sigma _{c} + \sigma _
{\overline{c}}})$.
There is a clear evidence that more leading $\Lambda _{c} ^{+}$ 
are produced at high $x _{F}$ in $\Sigma ^{-}$, $p$ beams. In the $\pi ^{-}$ 
beam
no asymmetry is found.
No charge bias in 
the asymmetry due to the trigger hodoscope requirement has been observed
in studies of these data.
To lowest order QCD, charm and anticharm quarks are 
produced symmetrically in hadroproduction. Next to Leading Order
(NLO) introduces small (1\%) asymmetries in quark momenta due to
interference between contributing amplitudes. 
The observed asymmetry can be explained by a recombination of charm
quark antiquarks with the beam valence quarks or by different
processes like in the intrinsic charm and in the quark-gluon 
string model \cite{theo}.
\par
In the SELEX experiment the  
($\Sigma ^{-}$ and $\pi^{-}$) have a
valence quark (d) in common with ${\Lambda _{c}}^{+}$ and $D^{-}$ but not
the $ D^{0}$.   Only the $\Sigma^-$ has a valence quark in common with
the $D_s^-$.  
The proton beam has two valence quarks in common with  ${\Lambda _{c}}^{+}$
(u and d quarks) and one quark with $D^{-}$ (d quark).  The pattern of
large and small cross sections and asymmetries in SELEX matches well with 
expectations from leading particle behavior.
\par
The SELEX results show in more detail trends observed in earlier
data.  E769  has observed a $\Lambda _{c}$ asymmetry integrated 
over $x _{F} >0$
for a 250 GeV/$c$ proton beam and in the same experiment has measured a
D meson asymmetry for a 250 GeV/$c$ pion beam
\cite{E769}.
The WA89 experiment has studied charm production in a 340 GeV/$c$ $\Sigma ^{-}$
 beam. Considerable production asymmetry between $D ^{-}$,
$D ^{+}$ and ${\Lambda_{c} }^{+} $, ${\overline{\Lambda_{c}}}^{-}$ was 
observed \cite{WA89}.

\section{Measurements of the $\Lambda _{c}$ ,$D ^{0}$ and $D _{s}$ lifetimes}

The charm decay modes used to measure the lifetime were
$\Lambda _{c} ^{+}  \rightarrow p K ^{-}  \pi ^{+}$;
$D ^{0}  \rightarrow  K ^{-} \pi ^{+}$ and 
$K ^{-}  \pi ^{+} \pi ^{-}\pi ^{+} $; and
$D _{s}  \rightarrow  K ^{*}(892) K$ and  $\phi \pi $.
The charm event selection criteria yield $1458 \pm 53 $ ,
$10090 \pm 131 $ and $918 \pm 53$ events after background subtraction 
for $\Lambda _{c} ^{+}$, $D ^{0}$ and $D _{s}$ respectively.
Candidate events have masses within 2.5$\sigma_M$ ( mass
resolution 8 $\rm{MeV/c}^{2}$) of their nominal charm mass value.
For the $D _{s}$ sample the resonance mass window for the 
$K^{*}(892)$ ($\phi$) was 
$892 \pm 70$  $\rm{MeV/c}^{2}$ ($1020 \pm 10$ $\rm{MeV/c}^{2}$).  
The yields for the $K^{*}(892)K$ and $\phi \pi$ channels 
are $395 \pm 30$ and $368 \pm 18$ signal events respectively.  
\par
$\pi /K$ misidentification causes a
reflection of $D^{*}$ and $D^{+}$ under the $D_{s}$ peak.  
The combination of the RICH Kaon identification
and the $\phi$ kinematics greatly reduces particle ID confusion
in the $\phi\pi$ channel.  For both samples  
we limit the maximum kaon momentum to 160 GeV/$c$ to reduce misidentification.  
We have studied the remaining
contamination by taking all $D_s$ candidates and computing the 
$D^{\pm}$ invariant mass obtained by 
replacing the $\rm{K}^{\pm}$ mass by the pion mass to evaluate a
$D_s^{\pm}$ candidate as a $\rm{D}^{\pm}$.  We label
any KK$\pi$ event having a pseudo-$D^{\pm}$ mass in an
interval of 20 MeV/$c^{2}$ at the central value of 1867 MeV/$c^{2}$ 
as a misidentified $D^{\pm}$ meson.  Some of these are real $D_s$ events; 
others really are misidentified $D^{\pm}$.  We remove them
all, to eliminate an artificial lengthening of the $D_s$ lifetime.
After sideband background subtraction we find about 30 $\%$ of the signal 
in the $K K \pi$ and $K^{*}(892)K$ channels is removed by the 
$D^{\pm}$ reflection subtraction.  For the $\phi\pi$ channel it is 5\%.
The statistical significance for the signal, $S/ \sqrt {S + B }$, is 
 $9.2 \pm 0.3$ and  $14.3 \pm 0.5$ for $K^{*}(892)K$ and $\phi \pi$
respectively. (S (B) is the number of signal (background) events in 
the signal mass region).
Because the bin-smearing effects are small, we used a binned maximum 
likelihood fitting technique to determine the lifetimes.
The fit was applied to a reduced proper time distribution, 
$ t^{*} ={M(l- l_{min})/ p c }$ where $l_{min}= N \sigma _{l} $ with N=8;
M and $p$ are the reconstructed mass and momentum values for each event.
The acceptance as a function of $t^{*}$ is evaluated using the set of  
observed events. Each event is  re-anayzed 1000 times with
randomly chosen $L$ values, holding event topology and momenta fixed.
This technique preserves all the production and acceptance properties and 
correlation of the data without doing a simulation. 
 A rethrown event is accepted if it passes the same cuts as those 
applied to the data \cite{sasha}.
For $t^{*}$ distributions in the signal and mass sideband regions
as shown in Figs. 2 and 3 we make a simultaneus fit to both signal
and sideband distributions. For  
$\Lambda _{c} ^{+}$ and $D ^{0}$ the sideband $t^{*}$ distribution is 
represented with a background function of two exponentials times acceptance.
For the $D_{s}$ the sideband $t^{*}$ distribution itself is normalized to the 
number of background events under the signal.  The lifetimes from SELEX are:
$\Lambda _{c} ^{+}$: $198.1 \pm 7.0  \pm 5.6$ fs; 
$D ^{0}$: $407.9 \pm 6.0 \pm 4.3 $ fs;
and $D_{s}$: $477.2 \pm 33.0  $ fs ($K^{*}(892)K$),
and $474.0 \pm 21.0 $ fs ($\phi \pi$).
Because systematic errors due to reflection subtraction are not
a problem,  we combine the $D _{s}$ resonant modes to give 
$\tau _{D_{s}} = 475.6 \pm 17.5  $ fs. 
The uncertainties are statistical only, evaluated where -ln L increases by 0.5.

\section{SUMMARY}
The SELEX experiment explores charm hadroproduction in the large $x _{F}$ 
region using different beams. 
The SELEX experiment finds clear $\Lambda _{c} ^{+}$ asymmetry production at large $x_{F}$ in $\Sigma$ and $p$ beams. 
We report a preliminary lifetime measurement of 
$\tau(\Lambda _{c})=198.1 \pm 7.0 \pm 5.5$ fs,
 $\tau(D ^{0})=407.0 \pm 6.0 \pm 4.3$ fs.
The preliminary $D _{s}$ lifetime, using two independent resonant decay channels, $K(892)^* K$ and $\phi \pi$, is
$ 475.6 \pm 17.5 \pm 4.4$ fs.  
Using our measured $ \tau _{D^{0}} $ we find a ratio,
$\tau _{D_{s}}/{\tau_{D^{0}}}$, 3.3$\sigma$ from unity. 
\par
The authors are indebted to the staff of Fermi National Accelerator
Laboratory and for invaluable technical support from the stuffs of 
collaborating institutions. This project was supported in part by
Bundesministerium f\"ur Bildung, Wissenchaft, Forschung und Tecnologie,
Consejo Nacional de Ciencia y Tecnologia (CONACyT), Conselho Nacional de
Desenvolvimento Cient\'ifico e Tecn\'ologico, Fondo de Apoyo a la
Investigaci\' on (UASLP), Funda\c c\~ ao de Amparo \'a 
Pesquisa do Estado 
de S\~ ao Paulo (FAPESP), the Israel Science Fundation founded by the
Israel Academy of Sciences and Humanities, Istituto Nazionale di Fisica
Nucleare (INFN), the International Science Foundation (ISF), the National Science Foundation (Phy 9602178), NATO (grant CR6.941058-1360-94), the Russian
Academy of Science, the Russian Ministry of Science and Technological Research
board (T\" UBITAK),the U.S. Department of Energy 
(DOE grant DE-FG02-91ER40664 and
DOE contract number DE-AC02-76CHO3000), and the U.S.-Israel Binational
Science Foundation (BSF).


\begin{thebibliography}{9}
\bibitem{Frix97} S. Frixione, M. Mangano, P. Nason and G. Ridolfi,
                 ``Heavy quark production in Heavy Flavour II'', 
                 A.J. Buras and M. Lindner eds. (World Scientific,
                 Singapore 1997)
                 I.I. Bigi and N.G. Uraltsev, Z. Phys. C62 (1994) 623
\bibitem{spec}   J.S. Russ, \emph{ et al.}, \emph{in Proceedings of the
                 29th International Conference on High Energy Physics,}
                 1998, edited by A. Astbury \emph{ et al.} (Word Scientific,
                 Singapore, 1998) Vol II, p. 1259; hep-ex/9812031  
\bibitem{theo}   T. Gutierrez and R. Vogt hep-ph/9808213
                 G.H. Arakelyan hep-ph/9711276 
                 A.K.Likhoded
                 and S.R. Slabospitsky, Preprint of IHEP97-66 Protvino
                 1997, hep-ph/9710476
\bibitem{E769}   E769 - G.A. Alves et al., Phys. Rev. Lett 77 (1996) 2388
\bibitem{WA89}   WA89 - M.I. Adamovich et al., Eur. Phys. J. C8 (1999) 593;
                 O. Piskounova this Proceeding
                 E791 - E.M. Aitala et al., Phys Lett. B411 (1997) 230; ibid
                 B403 (1997) 185; ibid B371 (1996) 157 
\bibitem{sasha}  A.Y. Kushnirenko, 
                 Ph. D. Thesis, Carnegie
                 Mellon University, 2000
\end{thebibliography}
\end{document}